\documentclass[twocolumn,aps,prc,longbibliography,nofootinbib,floatfix]{revtex4-1}
\usepackage{graphicx,bm,xcolor, bbold}
\usepackage[normalem]{ulem}
\usepackage{hyperref}
\hypersetup{colorlinks, linkcolor={blue},citecolor={blue},urlcolor={blue}}
\usepackage{graphicx}
\usepackage{amsmath,amssymb,amsfonts}
\usepackage{bm}
\usepackage{color}
\graphicspath{fig}
\usepackage{epstopdf}
\usepackage{multirow}
\usepackage{scalerel}

\newcommand{\paral}{\stretchrel*{\parallel}{\perp}}
\newcommand{\Eq}[1]{Eq.~(\ref{#1})}
\newcommand{\vect}[1]{\bm{\mathrm{#1}}}
\newcommand{\calE}{\mathcal{E}}
\newcommand{\Jv}{\vect{J}}
\newcommand{\jv}{\vect{j}}
\newcommand{\kv}{\vect{k}}
\newcommand{\nablav}{\vect{\nabla}}
\newcommand{\pv}{\vect{p}}
\newcommand{\Pv}{\vect{P}}
\newcommand{\qv}{\vect{q}}
\newcommand{\Qv}{\vect{Q}}
\newcommand{\rv}{\vect{r}}
\newcommand{\rhov}{\vect{\rho}}
\newcommand{\Vv}{\vect{V}}
\newcommand{\vv}{\vect{v}}
\newcommand{\Vbar}{\bar{V}{}}
\newcommand{\Vvbar}{\bar{\vect{V}}{}}

\newcommand{\xvhat}{\hat{\vect{x}}}
\newcommand{\yvhat}{\hat{\vect{y}}}
\newcommand{\zvhat}{\hat{\vect{z}}}
\newcommand{\lambdavpar}{\vect{\lambda}_{\paral}}
\newcommand{\sumpbppar}{\sum_{p_b,\pv_{\paral}}}
\newcommand{\calH}{\mathcal{H}}
\newcommand{\calR}{\mathcal{R}}

\DeclareMathOperator{\Real}{Re}
\DeclareMathOperator{\Imag}{Im}
\renewcommand{\Re}{\Real}
\renewcommand{\Im}{\Imag}

\begin{document}
\title{Superfluid fraction in the slab phase of the inner crust of neutron stars}
\author{Giorgio Almirante}
\email{giorgio.almirante@ijclab.in2p3.fr}
\affiliation{Universit\'e Paris-Saclay, CNRS/IN2P3, IJCLab, 91405 Orsay, France}
\author{Michael Urban}
\email{michael.urban@ijclab.in2p3.fr}
\affiliation{Universit\'e Paris-Saclay, CNRS/IN2P3, IJCLab, 91405 Orsay, France}
\begin{abstract}
An analysis of the slab phase as it is expected in the innermost layer of neutron-star crusts is performed within the Hartree-Fock-Bogoliubov framework. We take the periodicity of the slabs into account using Bloch boundary conditions, in order to well describe the interplay between the band structure and superfluidity. We introduce a relative flow between the slabs and the surrounding neutron gas in a time-independent way. This induces a non-trivial phase of the complex order parameter, leading to a counterflow between neutrons inside and outside the slabs. With the resulting current, we compute the actual neutron superfluid fraction. For the latter our results are slightly larger than previous ones obtained in normal band theory, suggesting that normal band theory overestimates the entrainment effect.
\end{abstract}
\maketitle
%
\section{Introduction}
The inner crust of neutron stars is expected to be composed by clusters of neutrons and protons, surrounded by a gas of unbound neutrons, together with a background degenerate relativistic electron gas that ensures charge neutrality and $\beta$-equilibrium \cite{Chamel08}. The clusters probably form a periodic lattice due to the interplay between the short-range nuclear force and the long-range Coulomb interaction \cite{Martin15,Thi21}, while the surrounding neutron gas has densities at which pure neutron matter is superfluid. The superfluid component of the crust could have observable consequences for the hydrodynamical and thermodynamical properties \cite{Page12}. It is also the main source of uncertainty in determining quantities such as shear modes \cite{Tews17}. Moreover, there is the belief that the superfluid component is involved in the mechanism that produces pulsar glitches \cite{Prix02,Carter06A}. In order to compare these models with observations, it is necessary to know some microscopic features of the inner crust \cite{Antonelli22}, such as the neutron superfluid fraction. 

The crucial point in computing the actual superfluid fraction is the evaluation of the so-called ``entrainment'', that is a non-dissipative force between the superfluid component and the nuclear lattice \cite{Prix02}. In fact the entrainment concerns all kinds of two-components systems with at least one superfluid part. One mechanism to explain this effect is the Bragg scattering, in our case of dripped neutrons by the nuclear lattice \cite{Chamel12A}, which is the analog of conduction electron scattering in ordinary solids. In practice, this requires very complicated band-structure calculations for the neutrons \cite{Chamel05,Chamel06,Chamel12A,Kashiwaba19,Sekizawa22}. The results of these calculations indicate that the entrainment can be very strong, reducing drastically the superfluid fraction. This is in contradiction with the observed glitches of certain pulsars \cite{Chamel13}, unless one gives up the common belief that only the crust is responsible for the glitches \cite{Andersson12}.

But one can also look at the entrainment from the perspective of superfluid hydrodynamics. In Ref.~\cite{Martin16}, the superfluid fraction was computed assuming an irrotational flow in a schematic density profile with simple boundary conditions between the clusters and the neutron gas. The results were, however, not in agreement with those obtained previously with the Bragg scattering approach. The entrainment obtained in the hydrodynamical approach is much weaker, and the corresponding superfluid fraction much larger, which if it was true would help to understand the observed Vela glitches.

Each of the two approaches has some shortcomings. On the one hand, the hydrodynamical approach \cite{Martin16} implicitly assumes that Cooper pairs are small compared to the spacing of the periodic lattice and to the size of the clusters, and unfortunately this is not true in the inner crust of neutron stars. On the other hand, band theory calculations \cite{Chamel05,Chamel06,Chamel12A}, even if pairing is added in the BCS\footnote{In this context, the BCS (Bardeen-Cooper-Schrieffer) approximation means that only diagonal matrix elements of the gap in the Hartree-Fock basis are included.} approximation \cite{Chamel10}, are missing the dynamics of the superfluid order parameter and therefore cannot reproduce superfluid hydrodynamics even when it should be valid, namely in the case of very strong pairing.

To reconcile the two approaches and solve this puzzle, it seems therefore necessary to go one step further, which is the Hartree-Fock-Bogoliubov (HFB) theory. Provided the periodicity of the system is taken into account (in contrast to the Wigner-Seitz approximation \cite{Pastore11} where only a single cell is considered), the HFB approach and its time-dependent extension (TDHFB) should include the full information of the band structure. Furthermore, unlike the BCS approximation, they should also be able to correctly reproduce the hydrodynamical behavior of the Cooper pairs in the limit of very strong pairing, as it was discussed in the context of cold atoms \cite{Grasso05,Tonini06}. The fact that the full HFB theory is needed and not only the simpler BCS approximation was demonstrated in the case of a toy model in Ref.~\cite{Minami22}. 

Therefore, in the present work, we address this problem by performing HFB calculations, at this time only in the slab (``lasagna'') phase. This kind of approach has recently been developed in Ref.~\cite{Yoshimura23} in a time-dependent framework. However, when the protons are accelerated as in Ref.~\cite{Yoshimura23}, we suspect that the entrained neutrons will always stay behind. Here we will therefore take a different approach. Namely, we perform static calculations including a relative flow between the slabs and the superfluid component in a stationary way. This is possible since we treat neutrons as superfluid but protons as normal, implying that if there is a flow, the state of our system will depend only on the relative velocity between the slab and the superfluid component (and not on two velocities as it would be the case if also protons were superfluid). Hence, a simple Galilean transformation is sufficient to retrieve a spatially periodic situation in spite of the flow. Another more technical difference between our study and Ref.~\cite{Yoshimura23} is the pairing interaction. We will use a momentum-dependent interaction that results in the correct density dependence of the gap.
In addition, we will study the slab phase both in physical and non-physical conditions, to compare with previous results and also to illuminate some interesting features that we can get with this kind of formalism.

In Sec.~\ref{sec:hamiltonian}, we present the interactions we use to construct the HFB matrix. In Sec.~\ref{sec:flow}, the formalism for two-fluid hydrodynamics and the inclusion of a relative flow are discussed. In Sec.~\ref{sec:results}, results without and with superfluid flow are shown and discussed for both unphysical and physical conditions. Section \ref{sec:conclusion} contains the conclusions and perspectives, and further details are given in the Appendix.

\section{Hamiltonian} \label{sec:hamiltonian}
Our aim is to compute the properties of the inner crust of neutron stars, which from a microscopic point of view consists of a system of neutrons and protons arranged in a periodic lattice. In order to do this we perform HFB calculations for neutrons and Hartree-Fock (HF) ones for protons (our system contains also electrons but they are fixed considering their distribution constant and requiring charge neutrality). These calculations require a mean-field Hamiltonian for both species and a pairing field for the neutrons, in the following we define these quantities.

The computational effort for this kind of calculations is quite big. Thus in this work we will focus only on the 1D periodic case (slabs), and postpone the 2D (rods) and 3D (crystal) cases to future work. In this section we keep the formalism general but the applications are performed with $L$-periodicity only in $x$ direction, details are given in Appendix \ref{appA}.

\subsection{Mean field}
For the mean-field we rely on a Skyrme energy-density functional, namely
%
%
\begin{align}
    \calE&_{\text{Skyrme}} = \nonumber\\
    &\int_V d^3 r \bigg[
    \frac{\hbar^2}{2m} \tau +
    \frac{t_0}{4}((2+x_0)\rho^2 + (2x_0+1)(\rho_n^2+\rho_p^2))\nonumber \\& +
    \frac{t_3}{24}((2+x_3)\rho^{\sigma+2} +
    (2x_3+1)\rho^\sigma(\rho_n^2+\rho_p^2))\nonumber\\ &+
    \frac{1}{8} (t_2(2+x_2)+t_1(2+x_1)) (\tau\rho-\jv^2)\nonumber\\& +
    \frac{1}{8} (t_2(2x_2+1)-t_1(2x_1+1))
    (\tau_n\rho_n+\tau_p\rho_p-\jv_n^2-\jv_p^2) \nonumber\\ &+
    \frac{1}{32} (t_2(2+x_2)-3t_1(2+x_1)) (\nablav\rho)^2 \nonumber\\ &+
    \frac{1}{32} (t_2(2x_2+1)+3t_1(2x_1+1))
    ((\nablav\rho_n)^2+(\nablav\rho_p)^2)\bigg]\,,
\end{align}
where $\rho = \rho_n+\rho_p$, $\tau = \tau_n+\tau_p$ and $\jv = \jv_n+\jv_p$ are total number density, kinetic energy density and momentum density, respectively. For each species $q = n,p$ these densities are defined as (we drop the index $q$ to avoid unnecessarily laborious notation, and summation over momentum should be understood as $\sum_{\kv} = \int d^3k/(2\pi)^3$)
\begin{gather}
    \rho(\rv)=2\sum_{\kv\kv'} \rho_{\kv\kv'}
      e^{i(\kv-\kv')\cdot\rv}\,,\\
    \tau(\rv)=2\sum_{\kv\kv'} (\kv\cdot\kv')
      \rho_{\kv\kv'} e^{i(\kv-\kv')\cdot\rv}\,,\\
    \jv(\rv)=\sum_{\kv\kv'} (\kv+\kv')
      \rho_{\kv\kv'} e^{i(\kv-\kv')\cdot\rv}\,,
\end{gather}
where
\begin{equation} \label{dens}
    \rho_{\kv\kv'}=\langle c_{\kv' \uparrow}^\dagger c_{\kv\uparrow}\rangle
    =\langle c_{\kv' \downarrow}^\dagger c_{\kv\downarrow}\rangle
\end{equation}
is the density matrix (we assume that there is no spin polarisation or spin current). For simplicity, we neglect the spin-orbit term. Concerning the parametrization, we use the SLy4 one, values for the coefficients and details can be found in \cite{Chabanat97}.

For the protons we include the Coulomb energy as follows
\begin{multline}
    \calE_{\text{Coul}} = \int_V d^3 r \bigg[
    \frac{1}{2}(\rho_p(\rv)-\bar{\rho}_e)V_C(\rv) \\
    -\frac{3e^2}{4}\Big(\frac{3}{\pi}\Big)^{1/3}
    (\rho_p(\rv))^{4/3}\bigg]\,,
\end{multline}
where the second term is the exchange term in the Slater approximation. The Coulomb potential $V_C$ is computed solving the Poisson equation
\begin{equation}
    \nablav^2 V_C(\rv) = 
    -4 \pi e^2 (\rho_p(\rv)-\bar{\rho}_e)\,,
\end{equation}
with $\bar{\rho}_e=\frac{1}{L}\int_0^L dx\,\rho_p(x)$ the electron density that compensates the charge of the protons.

Taking the functional derivatives of the energy density, one gets the mean-field Hamiltonian. Denoting
\begin{align}
    \frac{\hbar^2}{2m^*_q(\rv)} &= \frac{\delta\calE}{\delta\tau_q(\rv)}
    \,,\\
    U_q(\rv) &= \frac{\delta\calE}{\delta\rho_q(\rv)}
    \,,\\
    \Jv_q(\rv) &= -\frac{1}{2}\frac{\delta\calE}{\delta\jv_q(\rv)}\,,
\end{align}
and taking their Fourier transforms, the mean-field Hamiltonian for each species reads in momentum space
\begin{equation}
\label{eq:hmeanfield}
     h_{\kv\kv'} =
     \kv\cdot\kv'\Big(\frac{\hbar^2}{2m^*}\Big)_{\kv-\kv'} \hspace{-1mm}+
     U_{\kv-\kv'}
     - (\kv+\kv') \cdot \Jv_{\kv-\kv'}\,.
\end{equation}
%
\subsection{Pairing field}
For the pairing field we use a non-local interaction written in a separable form, namely
\begin{equation}
    V_{\kv_1\kv_2\kv_4\kv_3}^{\text{pair}} = -g
    \,
    f\Big(\frac{|\kv_1+\kv_2|}{2}\Big)
    f\Big(\frac{|\kv_3+\kv_4|}{2}\Big)
    \delta_{\kv_1-\kv_2,\kv_3-\kv_4}\,,
\end{equation}
where the matrix element of the pairing potential in the $^1S_0$ channel has to be understood as
\begin{equation}
    V_{\kv_1\kv_2\kv_3\kv_4}^{\text{pair}} =
    \langle \kv_1\!\!\uparrow\, -\kv_2\!\!\downarrow
    |V^{\text{pair}}| \kv_3\!\!\uparrow\, -\kv_4\!\!\downarrow \rangle\,.
\end{equation}
The minus signs in front of $\kv_2$ and $\kv_4$ have been introduced for convenience in accordance with those appearing in the Bogoliubov transformation for the $\downarrow$ particles.
The form factors $f(k)$ are taken to be Gaussians 
\begin{equation}
    f(k)=e^{-k^2/k_0^2}\,.
\end{equation}
The coupling constant $g$ and the Gaussian width $k_0$ had been fitted on the $V_{\text{low-}k}$ interaction in \cite{Martin14} with the result
\begin{equation}
    g = 853\,\text{MeV}\,\text{fm}^{3}
    \,;\quad
    k_0 = 1.365 \,\text{fm}^{-1}\,.
\end{equation}
With this interaction, the pairing field will be non-local too, and in momentum space it will read
\begin{equation} \label{eq:gapk}
    \Delta_{\kv\kv'} = g \hspace{1mm}
    f\Big(\frac{|\kv+\kv'|}{2}\Big) \sum_{\pv\pv'}
    f\Big(\frac{|\pv+\pv'|}{2}\Big) \kappa_{\pv\pv'}
    \delta_{\kv-\kv',\pv-\pv'}\,,
\end{equation}
where
\begin{equation} \label{eq:andens}
    \kappa_{\pv\pv'}=\langle c_{-\pv'\downarrow}
    c_{\pv\uparrow}\rangle\,,
\end{equation}
is the anomalous density matrix.\\
As it can be seen, the gap depends on the relative and the center-of-mass (c.o.m.) momentum separately.
\section{Flow} \label{sec:flow}
\subsection{Andreev-Bashkin matrix}
In order to study the flow in systems in which the superfluid phase and the normal one coexist, one can apply the formalism developed by Andreev and Bashkin \cite{Andreev75}. For the particle currents one has
\begin{alignat}{3}
\label{AB1}
    \rhov_n &=(\rho_n-\rho_{nn}-\rho_{np})\vv_N &+&
    \rho_{nn} \Vv_n &+& \rho_{np} \Vv_p\,,\\
\label{AB2}
    \rhov_p &=(\rho_p-\rho_{pp}-\rho_{pn})\vv_N &+&
    \rho_{pn} \Vv_n &+& \rho_{pp} \Vv_p \,,
\end{alignat}
where $\rho_n$ and $\rho_p$ are the number densities of neutrons and protons respectively. The other coefficients $\rho_{qq'}$
are the so-called Andreev-Bashkin matrix, an extension of the concept of superfluid density in the case of two fluids. The velocity of the normal part is denoted $\vv_N$, while $\Vv_n$ and $\Vv_p$ are the ``superfluid velocities'' of the superfluid component of neutrons and protons. These two last velocities are not velocities in the sense of matter displacement rate, in fact they are average momenta per unit mass, as pointed out in \cite{Prix04}. These velocities are defined as \cite{Andreev75}
\begin{equation} \label{eq:Vphase}
    \Vv_q=\frac{\hbar}{2m}\nablav\phi_q\,,
\end{equation}
where $\phi$ is the phase of the pairing field.
We are dealing with superfluid neutrons but only normal protons, Eqs.~(\ref{AB1}), (\ref{AB2}) are thus simplified because $\rho_{nn}$ is the only non-zero Andreev-Bashkin coefficient:
\begin{align} \label{AB3}
    \rhov_n &=(\rho_n-\rho_{nn})\vv_N + \rho_{nn} \Vv_n\,,\\
    \rhov_p &=\rho_p \vv_N\,.
\end{align}
This can be understood also in terms of Galilean invariance. Since protons are not superfluid, their current must be due to the only rigid motion of protons at the normal velocity. For the neutrons one has instead a superposition of the motion of the normal part and of the superfluid, and $\rho_{nn}$ can be identified with the neutron superfluid density, such that one can write
\begin{align} \label{AB4}
    \rhov_n &=(\rho_n-\rho_S)\vv_N +
    \rho_S \Vv_n\,,\\
    \rhov_p &=\rho_p \vv_N\,.
\end{align}
%
\subsection{Periodic inhomogeneities and Galilean transformation} \label{ssec:flowA}
It must be noticed that the relations above hold for homogeneous systems, but we are
considering a periodic structure of infinite slabs parallel to the $yz$ 
plane, with a period $L$ in $x$ direction.
We will apply these relations after averaging over one period. This is also a conceptual necessity, $\rho_S$ being a quantity that makes sense only in average. For other quantities such as densities and currents, we will indicate the cell average by a bar.\\
The neutron superfluid velocity component in the periodic dimension $x$ will be defined at a ``coarse-grained'' scale as follows:
\begin{equation} \label{eq:supvel}
    \Vbar_n = \int_0^L \frac{dx}{L}
    \frac{\hbar}{2m}\frac{\partial\phi}{\partial x}
    =
    \frac{\hbar}{2m}\frac{\phi(L)-\phi(0)}{L}\,.
\end{equation}
This means that, if there is a slow superfluid flow through the slabs, such that $\phi(L)-\phi(0)$ is not a multiple of $2\pi$, the gap is no longer periodic. This makes the HFB calculation very difficult, because we cannot use band theory any more.
However, we can avoid this difficulty by making a Galilean transformation to another frame, 
in which the slabs move, while the superfluid neutrons are at rest (in the sense $\Vbar_n=0$).
Let us explain this in some detail. We denote $S$ the frame
in which we do our calculations, while $S'$ is the rest frame of the slabs, moving 
with some velocity $\vv$ with respect to $S$. Since the protons are normal fluid and move together 
with the slabs, we conclude that $\vv = \vv_N$.

To simplify the discussion, let us start with the case of a single particle (e.g., neutron) moving in 
some potential $V$ (e.g., the mean field generated by the slabs). As discussed in the problem of \S 17 of \cite{LandauLifshitz3},
a single-particle wave function $\psi(\rv,t)$ in $S$ is related to the corresponding wave function 
$\psi'(\rv',t)$ in $S'$ by
\begin{equation}
    \psi(\rv,t)=\psi'(\rv-\vv t,t)e^{\frac{i}{\hbar}(m\vv\cdot\rv-\frac{1}{2}mv^2 t)}\,.
\end{equation}
We assume that the Hamiltonian $H'$ in $S'$ does not depend on time, while the Hamiltonian $H$ in $S$ 
is time dependent because the potential $V(\rv,t) = V'(\rv-\vv t)$ is moving. But at $t=0$, the coordinates $\rv$ 
and $\rv'$ coincide and therefore we have $V(\rv,t=0)=V'(\rv)$ and $H(t=0) = H'$. Taking a snapshot at $t=0$, we get
\begin{equation}
    \psi(\rv,0)=\psi'(\rv,0)e^{\frac{i}{\hbar}m\vv\cdot\rv}\,.\label{eq:snapshot}
\end{equation}
From now on we will always assume $t=0$ and drop the time argument.
Applying $H(=H')$ on both sides of \Eq{eq:snapshot}, assuming that $\psi'$ is stationary in $S'$
(i.e., $H'\psi'=E'\psi'$), one finds
\begin{equation}
  \Big(\frac{(-i\hbar\nablav-m\vv)^2}{2m}+V(\rv)\Big)\psi(\rv)\\
    = E'\psi(\rv)\,,
\end{equation}
or
\begin{equation} \label{eq:GalileanH}
  (H-\pv\cdot\vv)\psi(\rv) = \Big(E'-\frac{1}{2}mv^2\Big)\psi(\rv)\,,
\end{equation}
where $\pv = -i\hbar\nablav$. The relation (\ref{eq:GalileanH}) can be generalized to the many-particle case, if $H$ denotes the many-particle Hamiltonian and $\psi(\rv)$, $\pv$, and $m$ are replaced, respectively, by the many-particle wave function $\psi(\rv_1,\rv_2,\dots)$, the total momentum operator $\Pv$, and the total mass $M$ of the system.

Since the anomalous density matrix $\kappa$ involves the product of two wave functions (or field operators), it is clear from Eqs.~(\ref{eq:snapshot}) and (\ref{eq:supvel}) that the phase $\phi$ and superfluid velocity $\Vvbar_n$ in frame $S$ are related to the corresponding quantities $\phi'$ and $\Vvbar'_n$ in the slab rest frame $S'$ by
\begin{equation}
    \phi(\rv) = \phi'(\rv)+\frac{2m\vv\cdot\rv}{\hbar}\quad\text{and}\quad\Vvbar_n = \Vvbar'_n+\vv\,.
\end{equation}
Hence, if our HFB calculation in frame $S$ is constrained to $L$-periodic quantities, including the gap $\Delta$ and the phase $\phi$, we have a to set $\vv=-\Vvbar_n'$. Notice that, unlike having a phase $\Vvbar_n\cdot \rv$, adding the term $\pv\cdot\Vvbar'_n$ to the Hamiltonian does not destroy the periodicity.

In summary, we replace in the HFB equations the mean-field Hamiltonian (\ref{eq:hmeanfield}) by
\begin{equation}
\label{eq:Hpv}
     h_{\kv\kv'} =
     \kv\cdot\kv'\Big(\!\frac{\hbar^2}{2m^*}\!\Big)_{\!\kv-\kv'}\!
     + U_{\kv-\kv'}
     - (\kv+\kv') \cdot \Jv_{\kv-\kv'}
     - \hbar\kv\cdot\vv \delta_{\kv\kv'}.
\end{equation}
Due to the last term, there will be non-vanishing currents $\jv_q$ and the gap will become complex but remain periodic, corresponding to the situation of bound protons and neutrons flowing with velocity $v=v_N$ in a fluid of neutrons, the latter having a superfluid part that carries no momentum. 
Notice that it is somewhat sloppy to say we are in the superfluid rest frame, since $\Vvbar_n$ is not a true velocity. The true velocity cannot be defined separately for the normal part and the superfluid one, but only for a certain particle species, thus for neutrons in general.
In this scenario, knowing current and density of neutrons, one can rearrange \Eq{AB4} using that $\Vbar_n=0$ and get the superfluid density as
\begin{equation} \label{supdens}
    \rho_S=\bar{\rho}_n-\frac{\bar{\rhov}_n}{v_N}\,.
\end{equation}
where $\bar{\rho}_n$ and $\bar{\rhov}_n$ are respectively neutron density and current averaged over a period.
\subsection{Currents}
Now the point is how to compute the currents of neutrons and protons. The particle currents describe the actual matter displacement, thus they are the currents appearing in the continuity equation
\begin{equation}
    \frac{\partial}{\partial t} \rho_q(\rv,t) +
    \nablav \cdot \rhov_q(\rv,t) = 0\,.
\end{equation}
As shown in \cite{Chamel19,Allard21} in the framework of TDHFB theory, the currents can be expressed in terms of the effective mass and the momentum densities as 
\begin{equation} \label{eq:curr}
  \rhov_q(\rv,t)=\frac{\hbar}{m_q^*(\rv,t)} \jv_q(\rv,t)
    -\frac{2}{\hbar}\Jv_q(\rv,t)\rho_q(\rv,t)\,.  
\end{equation}
This expression takes into account the misalignment between particle and momentum transport due to presence of an effective mass term in the density functional. Despite this misalignment, Galilean invariance implies that for the total currents the relation
\begin{equation}
    m\rhov_n+m\rhov_p = \hbar \jv_n+\hbar \jv_p
\end{equation}
must hold.

It has to be noticed that \Eq{eq:curr} arises already in the Time-Dependent Hartree-Fock (TDHF) framework \cite{Engel75}, it continues to hold also in TDHFB unless a violation of the continuity equation is induced by the non-local pairing interaction. We discuss this point in Appendix \ref{appB}.

In our case we deal with a stationary one-dimensional flow, thus the relations above become function of the only variable $x$. We have access to densities and momentum densitites, since we compute them anyway in order to update the Skyrme mean-field potential. In this way we can compute the currents and through \Eq{supdens} get access to the superfluid density.

\section{Results} \label{sec:results}
\subsection{Densities and pairing field}
\label{subsec:density-gap}
Before considering a superfluid flow, let us briefly discuss the static case, i.e., $\vv=0$ in \Eq{eq:Hpv}.
In our calculations, we will analyze nuclear matter under $\beta$-equilibrium. This condition implies that the chemical potentials satisfy
\begin{equation}
    \mu_n=\mu_p+\mu_e\,,
\end{equation}
where $\mu_n$ is fixed and $\mu_e$ is computed with the ultra-relativistic expression
\begin{equation}
    \mu_e = \hbar c (3\pi^2\bar{\rho}_e)^\frac{1}{3}\,.
\end{equation}
The electron density is determined requiring charge neutrality, and $\mu_p$ is readjusted after each HFB iteration to satisfy the $\beta$-equilibrium condition.

\begin{figure}
\centering%
\includegraphics{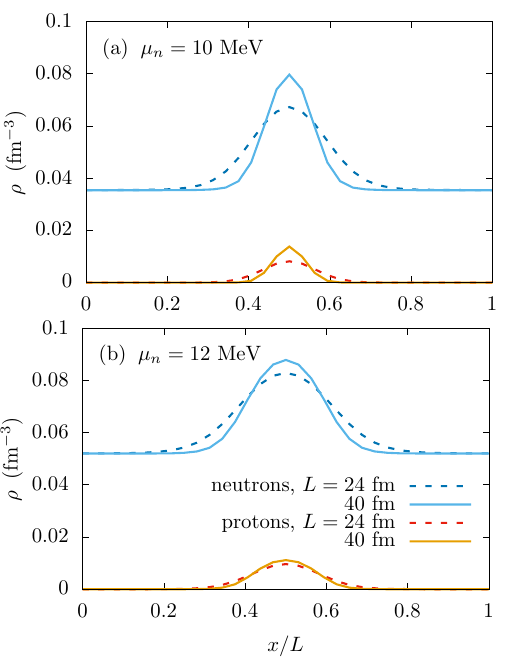}
\caption{Number densities $\rho$ for neutrons and protons at different chemical potentials $\mu_n$ and cell extensions $L$.}%
\label{fig:densities}%
\end{figure}

In principle, for given $\mu_n$, one should use the extension $L$ that minimizes the thermodynamic potential \cite{Martin15}, which is equivalent to minimizing the energy for given baryon density as done in \cite{Yoshimura23}. But the minimum is very flat and depends sensitively on details of the chosen interaction. In order to be able to compare with other calculations, e.g. \cite{Carter05}, we consider different values for the cell extension $L$ for a couple of choices for the neutron chemical potential $\mu_n$.
Some results for the density profiles are displayed in Fig.~\ref{fig:densities}. 
As it can be seen, protons (red and orange lines) are confined within the slab near the center of the cell, while the neutron density (blue lines) extends over the full cell. The neutron gas corresponds to the region where their density is constant. Increasing $L$ from 24 fm (dashed lines) to 40 fm (solid lines), we see that the density inside the slab increases, while the ratio between slab radius and cell extension (i.e., the volume fraction of the slab) decreases in such a way that the total (average) neutron and proton densities remain almost unchanged (see also Table \ref{betaequilibrium}). Notice that the gas density remains constant since in the homogeneous case the density is only a function of the chemical potential. Increasing $\mu_n$ from 10 MeV (upper panel) to 12 MeV (lower panel), we see that the neutron density increases both in the slab and in the gas, but the difference between the two decreases, the ratio between slab radius and cell extension increases and the system becomes progressively less inhomogeneous.

\begin{figure}
\includegraphics{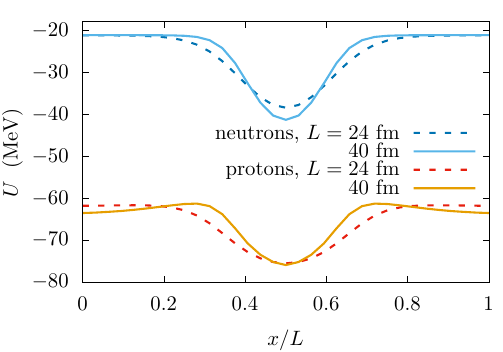}
\caption{Mean-field potential $U$ for neutrons and protons at different cell extensions $L$ for $\mu_n=12$ MeV.}%
\label{fig:potential}%
\end{figure}

For the sake of completeness, some mean-field potentials are shown in Fig.~\ref{fig:potential} for the case $\mu_n = 12$ MeV.
The behaviors with varying cell extension reflect those of the densities. The repulsive Coulomb interaction is visible in the proton potential outside the slab, especially in the case $L=40$ fm (orange curve).

\begin{figure}
\centering%
\includegraphics{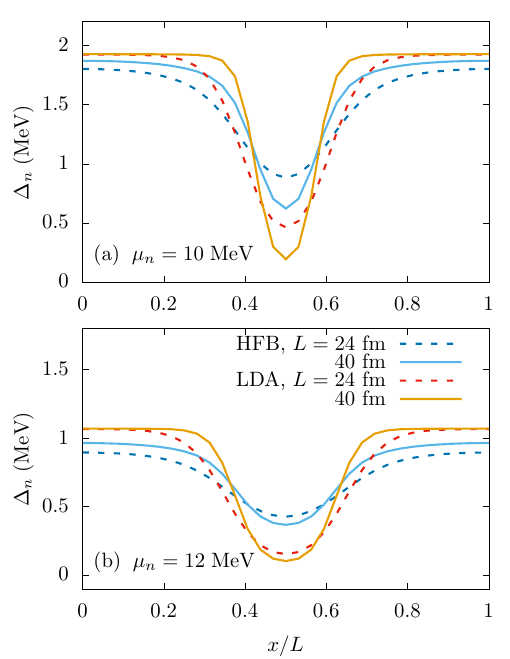}
\caption{Pairing gap $\Delta$ for neutrons at different chemical potentials $\mu_n$ and cell extensions $L$. The HFB gap is taken at the local Fermi momentum as in \Eq{eq:gapkf} while the LDA one is obtained by solving the BCS gap equation for uniform matter (see text for details).}
\label{gaps}
\end{figure}

Finally, let us discuss the spatial dependence of the pairing gap shown in Fig.~\ref{gaps}.
The HFB gap is computed by taking the expression (\ref{eq:gapkn}) in momentum space (i.e., \Eq{eq:gapk} constrained to a 1D periodic lattice) and performing the Fourier transform w.r.t. the c.o.m. momentum. Writing the gap in momentum space as
\begin{equation}
    \Delta_{n n'}(k_b,k_{\paral})= \frac{g}{(2\pi)^2} f_{n+n'}
    (k_b,k_{\paral}) F_{n-n'}\,, 
\end{equation}
where $n$ and $n'$ are the indices of the bands, $k_b$ is the Bloch momentum in $x$ direction, and $k_{\paral}$ is the momentum in the $yz$ plane (see Appendix \ref{appA} for details), the c.o.m. is related to the $n-n'$ part. This is due to our definition (\ref{eq:andens}) of the anomalous density, where the two particles carry, respectively, $-\kv'$ and $\kv$, and thus the c.o.m. momentum is $\kv-\kv'$. Given the separable form of the gap, one can write it as a function of the c.o.m. position and the relative momentum of the pair. Denoting $\Qv=(\frac{\pi}{L}(n+n')+k_b,k_y,k_z)$, $\nu=n-n'$, and
\begin{equation}
    \Delta_0(x)=\frac{g}{(2\pi)^2}\sum_{\nu}
    F_{\nu} \exp\Big(i\frac{2\pi}{L}\nu x\Big)\,,
\end{equation}
the non-local pairing gap can be written in Wigner (phase-space) representation as
\begin{equation} \label{eq:gapsep}
    \Delta(Q,x)= f(Q) \Delta_0(x).
\end{equation}
In Fig.~\ref{gaps}, the HFB gaps (blue lines) are evaluated with the form above for $Q$ equal to the local Fermi momentum, i.e.,
\begin{equation} \label{eq:gapkf}
    \Delta(k_F(x),x) = f\big(\sqrt[3]{3\pi^2\rho_n(x)}\big) \Delta_0(x)\,.
\end{equation}
This is more relevant than the gap at $Q=0$, and it is probably the quantity that can be best compared with momentum independent gaps obtained with contact interactions.

For comparison, we also display the gaps obtained with the Local-Density Approximation (LDA, red and orange lines), i.e., the values of $\Delta_{k_F}$ in uniform matter \cite{Martin14}, evaluated at the local density $\rho_n(x)$ with the same separable interaction as we are using in the HFB.
Namely, we replace $\Delta_0(x)$ in \Eq{eq:gapkf} by $\Delta_0^{\text{LDA}}(x)$ computed by solving the BCS gap equation
\begin{equation}
    1-g\sum_k\frac{f^2(k)}{2\sqrt{\big(\frac{k^2}{2m_n^*}-\mu_n^*\big)^2+\big(f(k)\Delta^{\text{LDA}}_0\big)^2}}=0\,,
\end{equation}
at each point $x$ with the local effective mass $m_n^*(x)$ and the effective chemical potential $\mu_n^*(x)$ chosen such as to reproduce the local density.\\
As it can be seen there is a strong reduction in the value of the gap inside the slab with respect to the gas. This is because in the slab the density is higher than in the gas, and in this range of chemical potential $\mu_n$ the neutron density is such that the gap decreases with increasing density (as seen in Fig.~1 of \cite{Martin14}, the $V_{\text{low-}k}$ gap in uniform matter is maximum around $k_F\approx 0.8$ fm$^{-1}$, i.e., at $\rho_n\approx 0.017$ fm$^{-3}$). For the same reason we find that the gap globally decreases when we increase $\mu_n$ from 10 (upper panel) to 12 MeV (lower panel).

Comparing the gaps in HFB (blue lines) and LDA (red and orange lines), we see that the HFB gaps are smaller than the LDA ones in the gas, while in the slabs it is the opposite. This behavior was already observed in Ref.~\cite{Chamel10} and was interpreted as proximity effect. In the gas, the agreement between HFB and LDA gets better if one increases the cell size $L$ from 24 (dashed lines) to 40 fm (solid lines), but it remains bad in the slabs. One way to see the interplay between the gap and the cell extension is to look at the ratio between the coherence length $\xi$ of the pair \cite{DeBlasio97} and the typical length scale on which the density varies in space, i.e., the cell extension $L$ but also the smaller length characterizing the thickness of the slab (which is still of the same order of magnitude as $L$). If $\xi$ is much smaller than the cell extension, the correlations between paired particles are less affected by the presence of inhomogeneities. This can also be seen as a hydrodynamical limit, since the pairs can be considered more and more as bosonic entities. The condition for the validity of the LDA can be written as
\begin{equation}\label{eq:coherencelength}
    \xi \ll L
    \quad\text{with}\quad
    \xi = \frac{\hbar^2 k_F}{\pi\Delta m^*}\,.
\end{equation}
This explains why the agreement between LDA and HFB in the gas is better for $\mu_n=10$ MeV (upper panel) than for 12 MeV (lower panel), and why the LDA is never a good approximation inside the slabs.

\subsection{Flow and superfluid fraction}
Now we will consider the properties of our system in the presence of a stationary flow. As discussed in Sec.~\ref{ssec:flowA}, we have at our disposal a snapshot of the slab flowing at constant velocity $v_N$ through a neutron gas whose superfluid part carries no momentum. In order to find the superfluid fraction $\rho_S$, the current should be linear in $v_N$ and therefore we limit ourselves to values of $v_N$ that are small enough so that neither the densities nor the pairing gap (except its phase) are changed. 

To justify why we can limit ourselves to the linear regime, let us estimate the order of magnitude of $v_N$ in the inner crust of neutron stars between glitches. Since we are interested in the relative flow between the neutron superfluid and the slab, we have to consider the difference $\delta\Omega=\omega_S-\omega_N$ between the rotation frequencies of the two. Using the expected $\delta\Omega\simeq10^{-2}-10^{-1}$ s$^{-1}$ for the Vela pulsar \cite{Ruderman76}, we get
\begin{equation}
    v_N = R_{\text{NS}}\times\delta\Omega \simeq
    4\times(10^{-7}-10^{-6})\;c \,,
\end{equation}
where $R_{\text{NS}}\simeq12$ km is the radius of the neutron star. For numerical reasons we cannot use such small values of the velocity. But notice that being in the linear regime for $v_N$ means that all the quantities that are velocity dependent (such as phase of the gap and currents) will be simply linear in the velocity, while $\rho_S$, densities, effective masses, etc., will be unchanged.

\begin{figure}
\centering%
\includegraphics{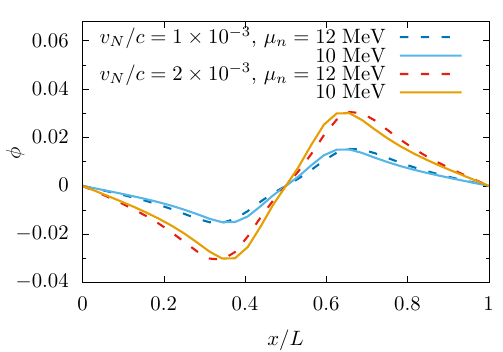}
\caption{Phase $\phi$ of the neutron pairing gap computed as in \Eq{eq:phi} at different chemical potentials $\mu_n$ and slab velocities $v_N$ for $L=24$\hspace{1mm}fm.}
\label{phase}
\end{figure}%

This allows us to study how a relative flow between the normal part and the superfluid influences the system. As said above, the superfluid carries no momentum, which follows from the definition we gave of the superfluid velocity in \Eq{eq:supvel}. The fact that the superfluid carries no momentum has to be understood at a ``coarse-grained'' level. However, at a microscopic scale within the cell this is no longer true because there is an entrainment between the slab and the superfluid. This can be seen in Fig.~\ref{phase}, where the phase of the gap is shown for different values of chemical potential $\mu_n$ and different velocities $v_N$. It has to be noticed that the phase of the gap depends only on the c.o.m. position of the pair, because of the separable form of the pairing gap, \Eq{eq:gapsep}, and the fact that its relative momentum part $f(Q)$ is real:  
\begin{equation} \label{eq:phi}
    \phi(x) = \arctan\Big(\frac{\Im\Delta_0(x)}{\Re\Delta_0(x)}\Big).
\end{equation}

First we see that the phase is $L$-periodic, which implies that $\Vbar_n=0$. But within the cell, the phase and hence also its derivative change. There are two points where the actual (microscopic) $V_n(x) = \partial\phi/\partial x$ is equal to zero and thus changes sign. Outside the slab, the superfluid velocity is opposite to the motion of the slab ($\partial\phi/\partial x<0$), while inside it is almost constant and it goes with the slab ($\partial\phi/\partial x>0$).

This behavior is similar to the one obtained in Ref.~\cite{Martin16} within a superfluid hydrodynamic approach. Assuming a schematic density profile, namely, constant density in the gas and in the slab with a sharp boundary between them, it was found that $V_n<0$ in the gas and $V_n>0$ in the slab. However, while in this simplistic approach the velocity was constant in each region with a discontinuity at the boundaries, there is no sharp boundary between the slab and the gas in the HFB calculation and therefore $\phi(x)$ and $V_n(x)$ vary continuously. Nevertheless, we see for instance that the distance between the two points where $V_n(x)=0$ is larger for $\mu_n=12$ MeV (dashed lines) than for $\mu_n=10$ MeV (solid lines), reflecting the different thickness of the slabs in the two cases (cf. Fig.~\ref{fig:densities}).

Doubling the slab velocity $v_N$, the phase of the gap, and hence the superfluid velocity, also doubles (red and orange vs. blue lines). This confirms that our system is indeed in the linear-response regime with respect to the slab velocity.
\begin{figure}
\centering%
\includegraphics{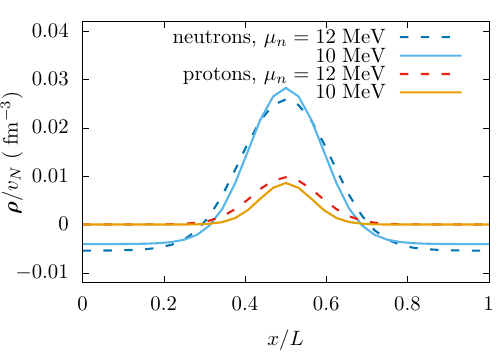}
\caption{Particle currents $\rhov$ for neutrons and protons computed as in \Eq{eq:curr} for different chemical potentials $\mu_n$ and $L=24$\hspace{1mm}fm at slab velocity $v_N/c=10^{-3}$.}
\label{currents}
\end{figure}%

Figure~\ref{currents} displays some results for the currents of neutrons and protons. The proton current is $\rhov_p(x)=v_N\rho_p(x)$ as expected, since protons are not superconducting. Concerning the neutrons, notice that in the case with no superfluidity the neutron current would be $\rhov_n(x)=v_N\rho_n(x)$, whereas the one we observe has the same spatial behavior but it is shifted down by a constant amount. This amount coincides with the (superfluid) neutron current in the frame in which the slab is at rest, and thus it has to be constant. A strong shift is a signal of a high neutron superfluid fraction since one can think that if all neutrons were superfluid, their current would be determined only by the superfluid velocity. In fact, one can see that there is indeed a counterflow outside the slab, as suggested by the behavior of the phase of the gap. It can be seen that with increasing chemical potential the neutron counterflow increases, as it was the case for the phase, in spite of the decreasing gap (see Fig.~\ref{gaps}). This is due to the reduction of the inhomogeneity, since in the homogeneous case at zero temperature all neutrons would be superfluid independently of the value of the gap (unless $\Delta=0$).

\begin{table}
  \caption{\label{betaequilibrium} 
    Results for baryon density $\rho_b$, average neutron gap at the local Fermi momentum $\bar{\Delta}_n$, superfluid fraction $\rho_S/\bar{\rho}_n$, Legget upper bound $\rho_{\text{leg}}/\bar{\rho}_n$ from \Eq{legget} and proton fractions $Y_p$ for different chemical potentials $\mu_n$ and cell extensions $L$ at $v_N/c=10^{-3}$.}
  \begin{ruledtabular}
    \begin{tabular}{ccccccc}
        $\mu_n$&$L$ &$\rho_b$&$\bar{\Delta}_n$&$\rho_S/\bar{\rho}_n$&$\rho_{\text{leg}}/\bar{\rho}_n$&$Y_p$\\
        (MeV)  &(fm)&(fm$^{-3}$)&(MeV)& & &(\%)\\\hline
        \multirow{3}{*}{10}&24&0.0441&1.52&0.9219&0.9528&3.14\\
                           &32&0.0436&1.56&0.9159&0.9439&3.13\\
                           &40&0.0430&1.60&0.9167&0.9414&3.13\\\hline
        \multirow{3}{*}{11}&24&0.0534&1.10&0.9325&0.9625&3.16\\
                           &32&0.0525&1.15&0.9318&0.9574&3.14\\
                           &40&0.0521&1.18&0.9334&0.9555&3.10\\\hline
        \multirow{3}{*}{12}&24&0.0626&0.73&0.9463&0.9730&3.23\\
                           &32&0.0622&0.75&0.9464&0.9698&3.23\\
                           &40&0.0615&0.79&0.9463&0.9677&3.16
    \end{tabular}
  \end{ruledtabular}
\end{table}

As discussed in Sec.~\ref{sec:flow}, having the neutron current $\rhov_n$, we can access the superfluid fraction. Our results are collected in Table \ref{betaequilibrium}, together with the baryon density $\rho_b$, the average neutron gap at the local Fermi momentum $\bar{\Delta}_n$ (i.e., the average over the cell of the quantity in \Eq{eq:gapkf}), the proton fraction $Y_p$, and the upper bound for the neutron superfluid density $\rho_\text{leg}$ derived by Leggett \cite{Leggett98} for inhomogeneous superfluids,
\begin{equation} \label{legget}
    \frac{1}{\rho_\text{leg}}=\int_0^L \frac{dx}{L} \frac{1}{\rho_n(x)}\,.
\end{equation}
As one can easily see from its definition, $\rho_{\text{leg}}$ is equal to $\rho_n$ in a uniform system and gets more and more reduced in the case of more pronounced inhomogeneities. Therefore, as one can see in the table, it increases with increasing neutron chemical potential $\mu_n$ and decreases with increasing cell extension $L$.

We find that the actual superfluid fraction $\rho_S/\bar{\rho}_n$ follows a similar trend. For fixed cell extension $L$, it increases with increasing neutron chemical potential $\mu_n$. This is mainly due to the progressive reduction of the inhomogeneity. In the dependence of the superfluid fraction on the cell extension at fixed chemical potential there is a competition that can be understood in terms of the coherence length of the pairs (cf. discussion about the HFB gap approaching the LDA in Sec.~\ref{subsec:density-gap}), assuming that the Leggett upper bound is reached in the hydrodynamical limit. It can be seen that with increasing $L$ the superfluid fraction becomes indeed closer to the Leggett upper bound.

\subsection{On the relation between pairing gap and superfluid fraction}
\begin{figure}
\centering%
\includegraphics{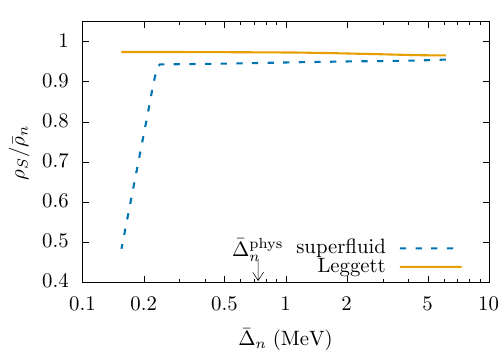}
\caption{Neutron superfluid fraction (blue dashes) and Leggett upper bound (orange line) as functions of the average gap obtained with an artificially scaled strength of the pairing interaction. The explored values for the pairing coupling constant are in the interval $g/g_{\text{phys}} \in [0.79,1.8]$, while the other parameters $\mu_n=12$ MeV, $L=24$ fm, and $v_N/c=10^{-3}$ are kept constant.
\label{supdensgap}}
\end{figure}

In the previous sections we studied the behaviors of pairing gap and superfluid fraction when changing the physical conditions of our system (chemical potential $\mu_n$ and cell extension $L$). In this way, the direct interplay between the two was hidden by the stronger effects due to the variation of the inhomogeneity. Here we perform an academic preliminary analysis to explore the dependence of the superfluid fraction on the pairing gap, changing by hand the value of the coupling constant $g$ in our pairing potential, such that the densities will be almost unchanged and only pairing gap and superfluid fraction will be affected. In Fig.~\ref{supdensgap} we display a comparison between the superfluid fraction and the corresponding Leggett upper bound.
%
%
The Leggett upper bound changes only at the $1\%$ level in the explored interval because the neutron density is only slightly affected by the variation of the gap. On the one hand, when we reduce the coupling constant by a factor of 0.79, the average pairing gap is reduced by almost one order of magnitude, while the superfluid fraction drops to its half. The onset of this reduction is rather sudden. The underlying physics to this behavior is related to the existence of a critical velocity $v_c$ where Cooper pairs start to be broken, the so-called Landau criterion (cf. \cite{LandauLifshitz9} for a one-component system and the recent work \cite{Allard23} for a homogeneous neutron-proton mixture). This effect is clearly no longer linear in $v_N$ and it sets in when $v_N$ and $v_c$ are of comparable magnitude. On the other hand, when we increase the coupling constant by a factor of 1.8, the average gap increases by one order of magnitude, but the superfluid fraction cannot increase very much because it is limited by the Leggett upper bound. Indeed, we see that with increasing gap, the superfluid fraction approaches the Leggett upper bound, which is what we suspected because the Cooper pairs behave more and more like pointlike bosons. A more detailed study of these phenomena goes beyond the scope of this work but we plan further investigations.

\subsection{Physical slab phase}
Until now our choices of chemical potential $\mu_n$ and cell extension $L$ were not matching the values expected for the slab phase in the inner crust of neutron stars. We have not explored the realistic parameters $\rho_b\simeq0.07-0.08$ fm$^{-3}$, $L\simeq 20-24$ fm \cite{Martin15,Thi21} in our previous analysis since the properties we discussed would have been less clear because of the much weaker inhomogeneities. According to \cite{Martin15}, for our values of $\mu_n$, the three-dimensional crystal lattice ($\mu_n\lesssim 12$ MeV) or the rod (``spaghetti'') phase ($\mu_n\simeq 12-13$ MeV) would be favored.

However, our method is also applicable in the physical slab phase since we studied the general 1D case for matter in $\beta$-equilibrium. Now we consider values for $\mu_n$ and $L$ that can be considered valid in the actual slab phase. In particular our aim is to compare our results with the ones obtained in Ref.~\cite{Carter05}. There the authors give definitions for the ``mobility tensor'' $K^{ij}$ and the density of ``free'' neutrons. The latter is a quantity whose definition relies on single-particle energy levels and it is not clear how to compute it in the HFB context. The former is instead a well defined quantity and relates current and momentum. Considering the slab phase, one can define the relevant mobility coefficients $K_{\paral}$ in the direction parallel to the slab and $K_\perp$ in the periodic one. They can be accessed from a macroscopic point of view as \cite{Carter06}
\begin{equation}
    K_{\paral}=\frac{\bar{\rho}_n}{m}\,,\quad
    K_\perp=\frac{\rho_S}{m}\,.
\end{equation}
%
\begin{table}
  \caption{\label{mobility}
    Results for baryon density $\rho_b$, average neutron gap at the local Fermi momentum $\bar{\Delta}_n$, ratio of mobility coefficients and proton fraction for different chemical potentials $\mu_n$ and cell extensions $L$ at velocity of the slab $v_N/c=1\times10^{-3}$.}
  \begin{ruledtabular}
    \begin{tabular}{cccccc}
      $\mu_n$&$L$&$\rho_b$&$\bar{\Delta}_n$&$K_{\paral}/K_\perp$&$Y_p$\\
      (MeV)&(fm)&(fm$^{-3}$)&(MeV)& &(\%)\\\hline
      \multirow{2}{*}{13}  &20&0.0723&0.42&1.0379&3.37\\
          &24&0.0720&0.43&1.0409&3.38\\\hline
      \multirow{2}{*}{13.5}&20&0.0768&0.31&1.0289&3.44\\
          &24&0.0766&0.32&1.0295&3.44
     \end{tabular}
  \end{ruledtabular}
\end{table}
\begin{table}
  \caption{\label{mobilityChamel}
    Results for baryon density $\rho_b$ and ratio of mobility coefficients for different chemical potentials $\mu_n$ and cell extension $L$ obtained in Ref.~\cite{Carter05} for the slab phase.}
  \begin{ruledtabular}
    \begin{tabular}{cccc}
        $\mu_n$&$L$&$\rho_b$&$K_{\paral}/K_\perp$\\
        (MeV)&(fm)&(fm$^{-3}$)& \\\hline
        31.70&23.71&0.0735&1.0698\\
        32.10&23.07&0.0749&1.0664\\
        32.79&22.23&0.0773&1.0605\\
        33.36&21.84&0.0792&1.0526
    \end{tabular}
  \end{ruledtabular}
\end{table}
%
The ratio between $K_{\paral}$ and $K_\perp$ is thus the inverse of the superfluid fraction we already computed. In Table~\ref{mobility} some of our results are shown, while in Table~\ref{mobilityChamel} there are the ones of Ref.~\cite{Carter05}. The difference in the chemical potential is due to the different models used. In Ref.~\cite{Carter05}, they consider a potential well with no effective mass and put the zero of the energy at the maximum value of the potential in the neutron gas (i.e., at $x=0$), such that, in order to have approximately the same density in the neutron gas, one has to put
\begin{equation}
    \mu_n^\text{HFB} \simeq \frac{m}{m^*_n} \mu_n^\text{well} + U_n(x=0)\,.
\end{equation}
In the cases we considered in Table~\ref{mobility}, we have $m/m^*_n\simeq 1.29-1.3$ and $-U_n(x=0)\simeq 27-28$ MeV, thus also the chemical potential ranges are compatible.

As it can be seen, we find a reduction in the perpendicular mobility coefficient but it is less pronounced than in \cite{Carter05}. Notice that in the physical slab phase the superfluid fraction is in both cases close to unity, thus the entrainment effect in this phase does probably not have any important astrophysical consequences. However, our results can be seen as another hint in the same direction as what was already pointed out in Refs.~\cite{Martin16,Watanabe17}, namely that the normal band theory of Ref.~\cite{Carter05} might systematically underestimate the superfluid fraction. If this was also true in the rod and crystal phases, where the entrainment is much stronger, it would have important implications for the understanding of glitch observations \cite{Andersson12,Chamel13}.

\section{Conclusions and perspectives}\label{sec:conclusion}
In this work we investigated the features of the periodic slab phase in $\beta$-equilibrium. This has been done performing HFB calculations with Bloch boundary conditions, making use of a Skyrme energy density functional for the mean-field and of a non-local separable potential fitted on $V_\text{low-k}$ for the pairing field. The latter point distinguishes our calculation from a similar approach in the literature \cite{Yoshimura23} which employs a contact interaction. In our approach, the suppression of the gap inside the slabs compared to the neutron gas is not due to an explicit density dependence of the coupling constant, but due to the momentum dependence of the pairing interaction.

First we studied the static situation, checking the general reduction of the pairing gap and determining the consequences of the presence of inhomogeneities on its behavior, comparing in particular with the LDA. Then we turned on a relative stationary flow between the slabs and the surrounding superfluid neutron gas, in a linear regime for the relative velocity. In this case we found that in the superfluid rest frame (in the sense $\Vbar_n = 0$) there is a counterflow of neutrons outside the slab, while the neutrons inside the slab move in the same direction as the slab, as is also clear from the periodicity of the phase of the gap. This means that in the rest frame of the slabs, the neutrons inside the slabs move more slowly than those of the gas. Moreover the entrainment has been quantitatively evaluated by computing the superfluid fraction, and a comparison with the corresponding Leggett upper bound has been performed. We found that in the slab phase with realistic parameters, the reduction of the superfluid fraction due to the entrainment effect is weaker than in previous calculations based on band-structure theory \cite{Chamel05}.

In contrast to \cite{Yoshimura23}, the way in which we include the relative flow is based on a simple Galilean transformation and does not require any time dependent calculation. In addition to its simplicity, this approach has the advantage that we are really describing a stationary state. It can be generalized to any geometry and number of spatial dimensions, but only if a single component is superfluid (i.e., no proton superconductivity). One of our objectives for the future is to extend our work to the phases that are periodic in two and three dimensions (rods, spheres). This is relevant since the differences we found with previous results \cite{Chamel05} could be enhanced in these geometries due to the stronger density variations which, as we saw, reduce the superfluid fraction. Clearly this can have astrophysical consequences, especially for the understanding of pulsar glitches \cite{Andersson12,Chamel13,Martin16,Watanabe17}.

When artificially decreasing the pairing interaction, we observed a strong reduction of the superfluid fraction. This hints toward an interesting effect, namely the breaking of Cooper pairs due to the superfluid flow beyond the Landau critical velocity, similar to the gapless superfluid phase discussed in \cite{Allard23} for the case of uniform matter (cf. also \cite{Urban08} for an analogous phase in ultracold atomic gases). This effect is beyond the linear regime and may also appear for the physical value of the pairing interaction, if the velocity of the flow is sufficiently fast, e.g., near a superfluid vortex. We plan further investigations on this subject. 
%
\acknowledgments
We wish to thank N. Chamel, V. Allard, N. Shchechilin, K. Sekizawa, and K. Yoshimura for interesting and helpful discussions and
detailed comparisons between our results.
%
\appendix
\section{HFB with Bloch boundary conditions in 1D} \label{appA}
We want to solve the HFB equations in a 1D periodic lattice. In order to do this, one can introduce the $L$-periodicity condition on the $x$-axis for the basic quantities, namely density and anomalous density, i.e.
\begin{gather}
\langle \psi^\dagger_\uparrow(\rv'+L\xvhat) \psi_\uparrow(\rv+L\xvhat) \rangle 
  = \langle \psi^\dagger_\uparrow(\rv')\psi_\uparrow(\rv) \rangle\,,
  \\
\langle \psi_\downarrow(\rv'+L\xvhat) \psi_\uparrow(\rv+L\xvhat) \rangle 
  = \langle \psi_\downarrow(\rv') \psi_\uparrow(\rv) \rangle\,,
  \label{perd}
\end{gather}
and require translational invariance in the $yz$-plane, i.e., $\forall \lambdavpar = a\yvhat+b\zvhat$ with $a,b\in \mathbb{R}$
\begin{gather}
\langle \psi^\dagger_\uparrow(\rv'+\lambdavpar) \psi_\uparrow(\rv+\lambdavpar) \rangle 
  = \langle \psi^\dagger_\uparrow(\rv') \psi_\uparrow(\rv) \rangle\,,
  \\
\langle \psi_\downarrow(\rv'+\lambdavpar) \psi_\uparrow(\rv+\lambdavpar) \rangle 
  = \langle \psi_\downarrow(\rv') \psi_\uparrow(\rv) \rangle\,.
  \label{homo}
\end{gather}
For the anomalous density matrix one can write
\begin{equation}
\langle c_{-\pv'\downarrow} c_{\pv\uparrow} \rangle
 = \int \! d^3 r\, d^3r' e^{-i\pv\cdot\rv}\, e^{i\pv'\cdot\rv'} \langle \psi_\downarrow(\rv') \psi_\uparrow(\rv) \rangle
   \,.
   \label{eq:ccavg}
\end{equation}
Performing a change in the integration variables $(\rv,\rv') \to (\rv+ L\xvhat + \lambdavpar,\rv' + L\xvhat + \lambdavpar)$ and using Eqs.~(\ref{perd}) and (\ref{homo}), one obtains
\begin{multline}
\langle c_{-\pv'\downarrow} c_{\pv\uparrow} \rangle 
  =\int\! d^3 r\, d^3 r'\, e^{-i\pv\cdot\rv}\, e^{i\pv'\cdot\rv'} \langle \psi_\downarrow(\rv') \psi_\uparrow(\rv) \rangle
  \\
  \times e^{-i(p_x-p'_x)L} e^{-i(\pv_{\paral}-\pv'_{\paral})\cdot\lambdavpar}\,.
  \label{eq:ccavgshifted}
\end{multline}
Combining Eqs.~(\ref{eq:ccavg}) and (\ref{eq:ccavgshifted}), one finds that the momentum labels must satisfy the following conditions: $p_x-p'_x=\frac{2\pi}{L}\nu$ with $\nu\in\mathbb{Z}$ and $\pv_{\paral}-\pv'_{\paral} = 0$ (where $\pv_{\paral} = p_y \yvhat+p_z\zvhat$ and analogously for $\pv'_{\paral}$). Hence, the momentum dependence of the matrix element can be written as
\begin{equation}
    \langle c_{-\pv'\downarrow} c_{\pv\uparrow} \rangle =
    \delta_{p_b,p'_b} \delta_{\pv_{\paral},\pv'_{\paral}}
    \langle c_{-n'\downarrow}(-p_b,-\pv_{\paral}) c_{n\uparrow}(p_b,\pv_{\paral}) \rangle\,,
    \end{equation}
where we rewrote the momentum component in $x$ direction as a sum of an integer multiple of $\frac{2\pi}{L}$ and the Bloch momentum defined in the first Brillouin zone, namely $p_x=\frac{2\pi}{L}n+p_b$ and $p'_x=\frac{2\pi}{L}n'+p'_b$, with $n,n'\in \mathbb{Z}$ and $p_b,p'_b\in (-\frac{\pi}{L},\frac{\pi}{L}]$.

For the normal density matrix one can proceed in a completely analogous way and gets
\begin{equation}
    \langle c^\dagger_{\pv'\uparrow} c_{\pv\uparrow} \rangle =
    \delta_{p_b,p'_b} \delta_{\pv_{\paral},\pv'_{\paral}}
    \langle c^\dagger_{n'\uparrow}(p_b,\pv_{\paral}) c_{n\uparrow}(p_b,\pv_{\paral}) \rangle\,.
\end{equation}

As a consequence of these relations, our HFB matrix is diagonal in $p_b$ and $\pv_{\paral}$. Moreover, in our HFB matrix there is no dependence on the orientation of the parallel momentum, i.e., it depends only on $p_{\paral}^2=p_y^2+p_z^2$.

Summarizing, for each couple $(p_b,p_{\paral})$ we have an HFB matrix defined in the integer momenta $n,n'$, namely
\begin{equation}\label{eq:HFBH}
 \calH=
 \begin{pmatrix}
     h-\mu & -\Delta \\
     -\Delta^\dagger & -\bar{h}+\mu
 \end{pmatrix} ,
\end{equation}
where $h$ is the mean-field Hamiltonian (including the term $-\pv\cdot \vv$), $\Delta$ is the pairing field and $\bar{h}_{\kv \kv'}=h_{-\kv' -\kv}$.
We diagonalize it, obtaining quasi-particles energies $E_\alpha(p_b,p_{\paral})$ and eigenvectors $(U^*_{\alpha n}(p_b,p_{\paral}),-V^*_{\alpha n}(p_b,p_{\paral}))$. In terms of these, the normal and anomalous density matrices are expressed as
\begin{gather}
    \rho_{n n'}(p_b,p_{\paral})=\sum_{E_\alpha>0} V_{n' \alpha}^*(p_b,p_{\paral}) V_{n \alpha}(p_b,p_{\paral})\,,\\
    \kappa_{nn'}(p_b,p_{\paral})=\sum_{E_\alpha>0} U_{n \alpha}^*(p_b,p_{\paral}) V_{n' \alpha}(p_b,p_{\paral})\,,
\end{gather}
with $E_\alpha = E_\alpha(p_b,p_{\paral})$. Then, using the short-hand notation
\begin{equation}
    \sum_{p_b\pv_{\paral}} = \frac{1}{2\pi^2} \int_0^\infty dp_{\paral}\,p_{\paral}\int_{\text{BZ}} dp_b\,,
\end{equation} 
one can compute the densities and pairing field as
\begin{equation} \label{eq:dens1D}
    \rho(x) = 2\sumpbppar
    \sum_{n n'} e^{i\frac{2\pi}{L}(n-n')x}\,\rho_{n n'}(p_b,p_{\paral})\,,
\end{equation}
\begin{multline}
    \tau(x) = 2\sumpbppar
    \sum_{n n'}e^{i\frac{2\pi}{L}(n-n')x}
    \rho_{n n'}(p_b,p_{\paral})\\
     \times\Big( \frac{4\pi^2}{L^2}nn'+\frac{2\pi}{L}(n+n')p_b+p_b^2+p_
     {\paral}^2\Big)\,,
\end{multline}
\begin{multline}
    \jv(x) = 2
    \sumpbppar
    \sum_{n n'}  e^{i\frac{2\pi}{L}(n-n')x}
    \rho_{n n'}(p_b,p_{\paral})\\
    \times\Big(\frac{\pi}{L}(n+n')+p_b\Big) \xvhat\,,
\end{multline}
\begin{multline} \label{eq:gapkn}
    \Delta_{n n'}(k_b,k_{\paral})= g f_{n+n'}(k_b,k_{\paral}) \sum_{m m'} \delta_{n-n',m-m'}\\
    \times\sumpbppar
    f_{m+m'}(p_b,p_{\paral}) \kappa_{mm'}(p_b,p_{\paral})\,,
\end{multline}
where
\begin{equation}
    f_{n+n'}(k_b,k_{\paral}) =
    \exp\bigg(-\frac{(\frac{\pi}{L}(n+n')+k_b)^2+k_{\paral}^2}{k_0^2}\bigg)\,.
\end{equation}

All of this is also true for the Hartree-Fock case, with the simplification that there are no anomalous density and pairing field. Thus, instead of the HFB matrix, only the mean field Hamiltonian $h_{nn'}(p_b,p_{\paral})$ needs to be diagonalized. Denoting its eigenvalues and eigenvectors $\epsilon_\alpha(p_b,p_{\paral})$ and $V_{\alpha n}(p_b,p_{\paral})$, the density matrix reads then
\begin{equation} \label{eq:densmatHF}
    \rho_{n n'}(p_b,p_{\paral})=\sum_{\epsilon_\alpha<\mu} V_{n' \alpha}^*(p_b,p_{\paral}) V_{n \alpha}(p_b,p_{\paral})\,.
\end{equation}

\section{Continuity equation in the HFB framework} \label{appB}
The HFB equations can be rearranged to get the continuity equation. Since we are performing static calculations we can write them in the compact form
\begin{equation}
    [\calH,\calR]=0\,,
\end{equation}
where the HFB matrix $\calH$ is given by \Eq{eq:HFBH} and the generalized density matrix $\calR$ is defined as
\begin{equation}
    \calR=
    \begin{pmatrix}
        \rho & \kappa \\
        \kappa^\dagger & 1-\bar{\rho}
    \end{pmatrix} .
\end{equation}
The commutator above is fulfilled because $\calR$ is constructed out of the eigenvectors of $\calH$. Now one can take the upper left part of the $2\times 2$ HFB equations, getting
\begin{equation}
\label{eq:continuity-commutator}
    [h,\rho]-\Delta\kappa^\dagger+\kappa\Delta^\dagger=0\,.
\end{equation}
In order to simplify the following discussion we will consider the mean-field Hamiltonian $h$ in the case without effective mass, namely
\begin{equation}
     h_{\kv\kv'} =
     \frac{\hbar^2}{2m} \kv^2 - \hbar\kv\cdot\vv
     + U_{\kv-\kv'}\,,
\end{equation}
such that current and momentum density coincide (up to a factor of $\hbar/m$) for each species. Rewriting also \Eq{eq:continuity-commutator} in momentum space, one gets
\begin{multline}
   \label{eq:continuity-momentumspace}
    0 = (\kv-\kv')\cdot\Big(\frac{\hbar^2}{2m} (\kv+\kv') - \hbar\vv\Big) \rho_{\kv\kv'}\\
    +\sum_{\pv} (U_{\kv-\pv}\rho_{\pv\kv'}-U_{\pv-\kv'}\rho_{\kv\pv})\\
    +\sum_{\pv}(\kappa_{\kv\pv}\Delta^*_{\kv'\pv} - \kappa^*_{\kv'\pv}\Delta_{\kv\pv})\,.
\end{multline}
Renaming $\kv' = \kv-\qv$, multiplying by $2ie^{i\qv\cdot\rv}/\hbar$ and summing over $\kv$ and $\qv$, one finds that the two terms in the second line of \Eq{eq:continuity-momentumspace} cancel each other, while the remaining terms become
\begin{equation}
    \nablav \cdot \Big(\frac{\hbar}{m}\jv(\rv)-\vv\rho(\rv)\Big) + \frac{2i}{\hbar}\sum_{\qv} e^{i\qv\cdot\rv} \chi_{\qv} = 0\,,
\end{equation}
where $\jv$ is the momentum density [in the general case with effective mass, $\jv$ has to be replaced by the current given in \Eq{eq:curr}], $\rho$ is the density, and
\begin{equation}
    \chi_{\qv} = \sum_{\kv,\pv} (\kappa_{\kv,\pv}\Delta^*_{\kv-\qv,\pv} - \kappa^*_{\kv-\qv,\pv}\Delta_{\kv,\pv})\,.
\end{equation}
To satisfy the continuity equation for slabs that move with constant velocity $\vv$, the term $\chi_{\qv}$ should vanish. Rewriting the above expression using a general form for the gap equation
\begin{equation}
    \Delta_{\kv\kv'} = \sum_{\pv\pv'} V\Big(\frac{\kv+\kv'}{2},\frac{\pv+\pv'}{2}\Big) \delta_{\kv-\kv',\pv-\pv'} \kappa_{\pv\pv'}\,,
\end{equation}
introducing the notation $2\Qv=\kv+\kv'-\pv-\pv'$ for the momentum transfer and taking into account the hermiticity of the potential, one gets
\begin{multline}
    \chi_{\qv} = \sum_{\kv,\pv,\Qv} \kappa_{\kv,\pv}\kappa^*_{\kv-\qv-\Qv,\pv-\Qv} \\
    \times\Big[ V\Big(\frac{\kv+\pv-\qv}{2}-\Qv,\frac{\kv+\pv-\qv}{2}\Big) \\ -  V\Big(\frac{\kv+\pv}{2}-\Qv,\frac{\kv+\pv}{2}\Big)\Big]\,.
\end{multline}
As it can be seen, the above expression is equal to zero if the pairing interaction is local, since then it depends only on the momentum transfer $Q$ [i.e., $V(\kv,\kv')=V(\kv-\kv')$]. This is, however, not true for our separable pairing potential, thus the continuity equation can be violated. Anyway we found that this violation is four orders of magnitude less than the scale on which the current varies, such that we can neglect it.
\section{Numerical details} \label{appC}
As shown in Appendix \ref{appA}, our problem is diagonal in the Bloch momentum $k_b$ and in the modulus of the parallel momentum $k_{\paral}$, and thus we deal with an HFB matrix in the integer momenta $n,n'$. Since we want to perform the self-consistent calculations on a machine, the problem has to be discretized. We choose to divide the cell extension $L$ such that $L=N\Delta x$. This naturally introduces a cutoff in the integer momentum
\begin{equation}
    \Lambda=\frac{2\pi}{L}N \quad {\Rightarrow}\quad n\in (-N/2,N/2]\,.
\end{equation}
As a consequence, the HFB matrix has dimension $2N\times2N$ (the HF matrix instead only $N\times N$). In this way, we can access the first $N$ bands for both neutrons and protons. In this work, we choose $N=32$.

For each combination $(k_b,k_{\paral})$, a diagonalization of both the HFB and the HF matrices in momentum space is performed, the relevant quantities to construct the matrices are computed in coordinate space and then transformed through Fast Fourier Transform (FFT).\\

For $k_b$ of neutrons, we take $N_b=30$ points in the first Brillouin zone. For $k_{\paral}$ of neutrons, we introduce a cutoff $\Lambda_{\paral}^{(n)}=4$ fm$^{-1}$, which is sufficiently large compared to the parameter $k_0$ in the pairing interaction, such that the pairing gap does not depend on it. In the interval $[0,\Lambda_{\paral}^{(n)})$, we use $N_{\paral}=476$ points. This number should be a multiple of the available number of cores of the machine since we perform in parallel the diagonalizations for each $k_{\paral}$. With these choices, the average grid spacings are approximately
\begin{equation}
    \Delta k_b = \frac{2\pi}{N_b}L \approx \frac{0.21}{L}\,,\quad
    \Delta k_{\paral}=\frac{\Lambda_{\paral}^{(n)}}{N_{\paral}}\approx 0.0084\,\text{fm}^{-1}.
\end{equation}
However, notice that we use Gauss-Legendre points for both the $k_b$ and $k_{\paral}$ integrations, except when the pairing gap at zero momentum becomes particularly small (below $0.7$ MeV). In that case, we use an equidistant integration grid for both momenta but taking twice as many ($N_{\paral} = 952$) points in $k_{\paral}$ direction, i.e., $\Delta k_{\paral}\approx 0.0042$ fm$^{-1}$.\\

For protons, we distinguish two cases. If they are strongly confined, only a few bands are occupied and they are flat. In this case ($\mu_n<13$ MeV), we replace the integration over the Brillouin zone by multiplying by $2\pi/L$ the integral over $k_{\paral}$. The latter is performed with equidistant points with cutoff $\Lambda_{\paral}^{(p)}=2$ fm$^{-1}$. For protons, the cutoff can be smaller than for neutrons because of the absence of pairing. With $N_{\paral}=476$, the corresponding spacing is $\Delta k_{\paral}\approx 0.0042$ fm$^{-1}$. For $\mu_n\ge13$ MeV, the proton bands are no longer completely flat and therefore we perform also the integration over the Brillouin zone, with equidistant points and the same spacing $\Delta k_b$ as for neutrons.\\

Since protons are in the normal phase, their distribution is a step function and this makes the numerical integration more difficult. In this sense, in order to improve the precision of the $k_{\paral}$ integration, we perform the following trick. Consider \Eq{eq:dens1D} in the case in which protons are strongly confined, expressing the density matrix $\rho_{n n'}$ as in \Eq{eq:densmatHF} and expliciting the theta function:
\begin{multline}
    \rho(x) = \frac{2}{(2\pi)^2} \frac{2\pi}{L} \int_0^\infty\! dk_{\paral} \,k_{\paral}\,
    \sum_{n n'} e^{i\frac{2\pi}{L}(n-n')x}\\
    \sum_\alpha V_{n' \alpha}^*(k_{\paral}) V_{n \alpha}(k_{\paral}) \Theta(\mu_p-\epsilon_\alpha(k_{\paral}))\,.
\end{multline}
Noticing that only the step function varies abruptly, while the eigenvectors $V_{n\alpha}(k_{\paral})$ depend continuously on $k_{\paral}$, one can approximate the integral over the parallel momentum as a sum over the sample points $k_{\paral i}$, employing the following prescription:
\begin{multline}
    \rho(x) = \frac{1}{\pi L} \sum_{i=0}^{N_{\paral}-1}
    \sum_{n n'} e^{i\frac{2\pi}{L}(n-n')x}
    \sum_\alpha V_{n' \alpha}^*(k_{\paral i}) V_{n \alpha}(k_{\paral i}) \\
    \times\int_{k_{\paral i}}^{k_{\paral i+1}} dk_{\paral}
    \,k_{\paral} \, \Theta(\mu_p-\epsilon_\alpha(k_{\paral}))\,.
    \label{eq:densitykparallelintegral}
\end{multline}
To approximate the integral in the second line, one can consider the small difference between the energy in $k_{\paral i}$ and in $k_{\paral}$ in the framework of perturbation theory, writing
\begin{equation}
    h=h_0+h_1 \quad\text{with}\quad h_1=\frac{\hbar^2 \delta k_{\paral}^2}{2m^*(x)}\,,
\end{equation}
where the perturbative parameter is $\delta k_{\paral}^2 = k_{\paral}^2 - k_{\paral i}^2$ for $k_{\paral}\in [k_{\paral i},k_{\paral i+1})$. The energy correction at the first order is given by 
\begin{multline}
    \delta\epsilon_\alpha(k_{\paral i},\delta k_{\paral}^2)
    = \delta k_{\paral}^2 \sum_{n n'} V_{n' \alpha}^*(k_{\paral i})
    \Big(\frac{\hbar^2}{2m^*}\Big)_{nn'}
    V_{n \alpha}(k_{\paral i})\\
    \equiv \delta k_{\paral}^2 \Big\langle\frac{\hbar^2}{2 m^*}\Big\rangle_\alpha\,.
\end{multline}
In this way the remaining parallel momentum integral in the second line of \Eq{eq:densitykparallelintegral} for the density becomes
\begin{equation}
    I_{\alpha i} = \int_{k_{\paral i}}^{k_{\paral i+1}} dk_{\paral}
    \,k_{\paral}\,
    \Theta\Big(\mu_p-\epsilon_\alpha(k_{\paral i})-\delta k_{\paral}^2 \Big\langle\frac{\hbar^2}{2 m^*}\Big\rangle_\alpha\Big)
    \,.  
\end{equation}
Now, defining $t=\delta k_{\paral}^2\langle\hbar^2/2 m^*\rangle_\alpha$ such that $t_{\text{max}}=(k_{\paral i+1}^2 - k_{\paral i}^2)\langle\hbar^2/2 m^*\rangle_\alpha$ and changing variable in the integral above one gets
\begin{align}
    I_{\alpha i} =& \Big\langle\frac{\hbar^2}{2 m^*}\Big\rangle_\alpha^{-1} \int_{0}^{t_{\text{max}}} \frac{dt}{2} \,
    \Theta(\mu_p-\epsilon_\alpha(k_{\paral i})-t)\nonumber\\
    =&\begin{cases}
    0 & \text{if~} \epsilon_\alpha(k_{\paral i})\ge\mu_p\,,\\
    (k_{\paral i+1}^2 - k_{\paral i}^2)/2 & \text{if~} \epsilon_\alpha(k_{\paral i})+t_{\text{max}}\le\mu_p\,,\\
    \frac{1}{2}\big\langle\frac{\hbar^2}{2 m^*}\big\rangle_\alpha^{-1}(\mu_p-\epsilon_\alpha(k_{\paral i})) 
      & \text{else}\,.
    \end{cases}
\end{align}
With this method one is able to recognize the subinterval in which the distribution drops to zero and to better treat the integration in this case, even before knowing the energy $\epsilon_\alpha(k_{\paral i+1})$ in the next point, which simplifies the organization of the code.

All the choices we discussed about integration points and cutoffs are justified by the fact that if we decrease the former or increase the latter the self-consistent calculations converge to the same results. Our convergence criterion is defined such that for each point in the cell
\begin{align}
    |\rho^{(m)} - \rho^{(m+1)}| &< |\rho^{(m+1)}|\times10^{-4},\nonumber\\
    |j^{(m)} - j^{(m+1)}| &< |j^{(m+1)}|\times10^{-4},\nonumber\\
    |\Delta_0^{(m)} - \Delta_0^{(m+1)}| &< |\Delta_0^{(m+1)}|\times10^{-4},
\end{align}
where $\rho^{(m)}$ is the result of the $m$-th iteration etc.

In order to speed up the convergence, we use the Broyden's modified method as discussed in \cite{Baran08} and already used in a framework similar to ours in \cite{Yoshimura23}. Our Broyden vector has dimension $7 N + 2$ and it is defined as ($U_n(x)$, $U_p(x)$, $\hbar^2/2m^*_n(x)$, $\hbar^2/2m^*_p(x)$, $J_n(x)$, $J_p(x)$, $F_{n-n'}$, $\mu_n$, $\mu_p$). We performed the method using the results of $M=3$ previous iterations and a mixing coefficient $\alpha=0.7$.
\bibliography{refs}
\end{document}